\documentclass[conference]{IEEEtran}
\IEEEoverridecommandlockouts
\usepackage{cite}
\usepackage{amsmath,amssymb,amsfonts}
\usepackage{algorithmic}
\usepackage{graphicx}
\usepackage{textcomp}
\usepackage{xcolor}
\usepackage{comment}
\usepackage{comment}
\usepackage[normalem]{ulem}
\usepackage{float}
\newcommand{\fixme}[1]{{\textcolor{red}{\em\bf{[FIXME: #1]}}}}

\def\BibTeX{{\rm B\kern-.05em{\sc i\kern-.025em b}\kern-.08em
    T\kern-.1667em\lower.7ex\hbox{E}\kern-.125emX}}
\begin{document}

\title{A Graph Convolutional Neural Network based Framework for Estimating Future Citations Count of Research Articles
}
\author{
\IEEEauthorblockN{Abdul Wahid}
\IEEEauthorblockA{Department of CSE\\
Indian Institute of Technology\\
(Indian School of Mines), Dhanbad, India\\
Email: abdul.cspg14@nitp.ac.in}\\  
\and
\IEEEauthorblockN{Rajesh Sharma}
\IEEEauthorblockA{Insitute of Computer Science\\
University of Tartu, Estonia\\
Email: rajesh.sharma@ut.ee}\\ 
\and
\IEEEauthorblockN{Chandra Sekhara Rao Annavarapu}
\IEEEauthorblockA{Department of CSE\\
Indian Institute of Technology\\
(Indian School of Mines), Dhanbad, India\\
Email: acsrao@iitism.ac.in}             
}
\maketitle

\begin{abstract}
Scientific publications play a vital role in the career of a researcher. However, some articles become more popular than others among the research community and subsequently drive future research directions. One of the indicative signs of popular articles is the number of citations an article receives.
The citation count, which is also the basis with various other metrics, such as the journal impact factor score, the $h$-index, is an essential measure for assessing a scientific paper's quality. In this work, we proposed a Graph Convolutional Network (GCN) based framework for estimating future research publication citations for both the short-term (1-year) and long-term (for 5-years and 10-years) duration. We have tested our proposed approach over the AMiner dataset, specifically on research articles from the computer science domain, consisting of more than 0.8 million articles.
\begin{comment}
\fixme{add??--in particular on research articles from computer science domain, consisting of more than XX articles and approximately ABC authors.} \sout{It has a comprehensive collection of scholarly records, with more than 2 million articles and approximately 1.8 billion authors' profiles from several publication databases.}
\fixme{next lines need to be upgraded as we need to report some results here also.}
\end{comment}
By exploring both conventional and graph-based features, we have compared machine learning algorithms (Linear Regression, Random Forest, XGBoost, and Deep Neural Networks) as baseline methods with our GCN-based approach, which outperforms baseline algorithms in terms of error rates and $R^2$ value, indicating the robustness of the model.

\end{abstract}

\begin{IEEEkeywords}
graph convolutional neural networks, citation prediction, deep learning, academic networks
\end{IEEEkeywords}

\section{Introduction}\label{sec:Intro}
\par
\begin{figure}[h]
    \centering
    \includegraphics[width=\linewidth,height=35cm,keepaspectratio]{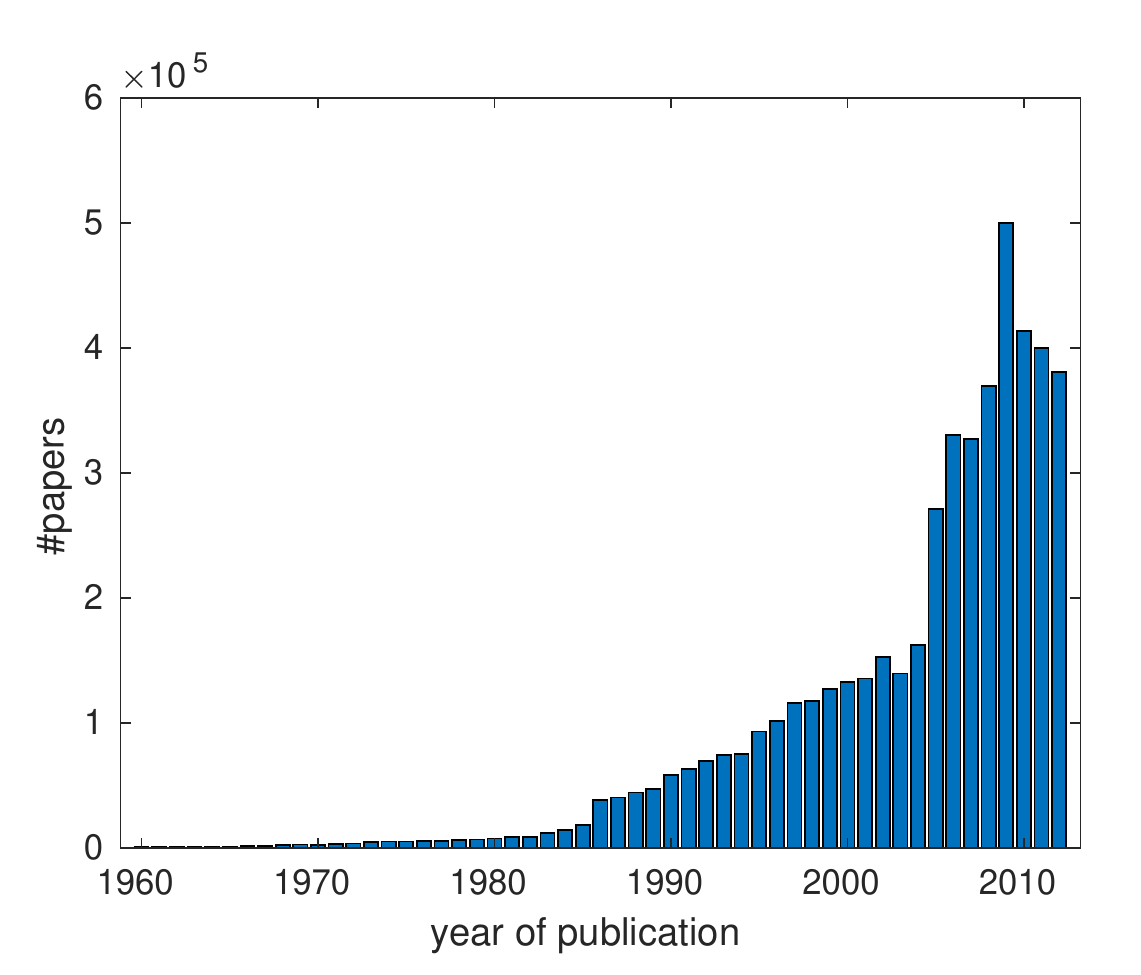}
    \caption{Number of papers published over the year.}
    \label{fig:fig1}
\end{figure}

 Scientific publications are vital and crucial for the development of the career of researchers. For example, well-conducted research published at a reputed venue enhances the reputation of a researcher. In the past decade, there has been a surge in the publications of research articles. For example, the statistics of research articles in the Computer Science domain from the AMiner dataset\footnote{https://www.aminer.cn/aminernetwork\\ This dataset is available from 1960 to 2012, however we considered it till December 2011 in our paper.} is shown in Figure \ref{fig:fig1}, which resembles this trend, where it can be observed that the number of articles being published is nearly three times compared to what has been published ten years ago.
\begin{comment}
\fixme{I have few issues with this FIGURE -- 1) it only shows till 2010 and we are in 2020? So it is very old information. 2) Also are you copy pasting this figure or you plotted it yourself ?}
\end{comment}
\par Due to the immense increase in the number of articles, the need to ascertain the qualities and impact of published articles is overwhelming. However, one of the ways to measure the scientific impact of the published articles is by citation count, which could be considered as one of the academia's strongest currencies. However, predicting the citations count could be challenging as identifying the factors influencing the citations of a research article is a non-trivial task. At the same time, the pre-identification of potentially significant papers will help scientists to invest in the research domains, which can potentially bring advancement in their career by projecting them as potential leaders and also attracting citations and grants.
\par In the past years, the task of citations counts prediction has gained considerable attention as the usage for the citations counts for evaluating the scientific impact has become more prominent \cite{ref20}. Various techniques like network science \cite{thelwall2018could}, machine learning \cite{ref21}, and mix of aforementioned techniques have been explored for predicting citations count \cite{ref8}. It has been shown that an article's early citations play a crucial role in predicting its long-run citation \cite{ref4}. In \cite{singh2017understanding}, it is found that early influential citations had an adverse effect on long-term scientific impact, maybe due to the attention-stealing, whereas early non-influential citations had a positive effect on the long-term scientific impact. In a different work, citations count prediction have used a generative model to reflect the observation that older papers generally receive higher citations \cite{newman2009first}. Recently in \cite{xiao2016modeling}, the authors introduced a point process model for predicting the long-term effect of individual publications based on early citations.
\par Briefly, past studies mainly rely on early citations of an article's impacts prediction. 
They primarily focused on the auto-correlation in the citations network of historical data.
The limitation of these predictive methods is that they rely heavily on historical citations. However, exploring the fundamental characteristics of the citations can reveal a new predictive approach that ignores the early citations. Compared to previous works, this paper contributes in the development of a new predictive model for the citations prediction by considering the following types of features:

\begin{comment}
\textcolor{blue}{In this paper, the citation count prediction problem is motivated by the following observations:
\begin{enumerate}
    \item There is a lack of citation count prediction methods to predict citations for both short-term and long-term duration simultaneously.
    \item To the best of our knowledge, this is the first time that the Graph Convolutional Network (GCN) is applied in the field of citation count prediction.
    \item The prediction accuracy of the CCP method is also a concern.
\end{enumerate}}
\end{comment}
\begin{itemize}
    \item \textbf{Article features:} This includes the title, the abstract, references, citation quality, popularity, and diversity of the article.
    \item \textbf{Author features:} This includes the number of authors in a paper, authors' affiliation, authors’ research interests, etc.
    \item \textbf{Venue features:} We also considered the venue features, like, the average citations and the h-index of the venue.
    \item \textbf{Network features:} We created a co-citation network to capture network features such as degree or in-degree. These features are representing the global influence in this citations count prediction problem.
\end{itemize}
We model the citation prediction as a regression learning task and exploited Graph Convolutional Network (GCN) \cite{kipf2016graph} to learn the prediction task. In recent years, graph convolutional networks have achieved significant success in various domains, such as in molecular structure analysis \cite{gilmer2017neural}
, community detection \cite{bruna2017community}
, link prediction \cite{chen2018gc}
, forecasting retweet count \cite{vijayan2018forecasting}
and image recognition. To the best of our knowledge, this is the first work that has explored Graph Convolutional Network (GCN) in the field of citation count prediction problem.
\par We applied our model on a real dataset consisting of more than 0.8 million articles taken from AMiner dataset\footnote{https://www.aminer.cn/aminernetwork}. We compared our GCN based framework with baseline methods (Linear Regression \cite{lovaglia1991predicting}, Random Forest \cite{ref37}, XGBoost \cite{ref36}, and Deep Neural Networks \cite{ref38}) where GCN achieved a MAPE score of 0.1072 and $R^2$ of 0.9157, which shows the effectiveness of the proposed framework.
\begin{comment}
\par \textcolor{blue}{The main contributions of this paper are summarized as follows:
\begin{enumerate}
    \item We have explored the features that reveal the fundamental characteristics of the citations. Some important features based on real-world observations with a citation and co-authorship network like citation quality, the first author’s citations count, and in-degree and out-degree in the network features are considered.
    \item A Graph Convolutional Network (GCN) is introduced first time in the citation count prediction problem to capture implicit relations between different features.
\end{enumerate}}
\end{comment}
\par The rest of the paper is organized in the following way —Section \ref{sec:background} reviews related work. Section \ref{sec:data} presents the problem definition, and a detailed discussion on the dataset which is used in our experiments. Section \ref{sec:method} covers methodology proposed in this paper, including a suitable measure of paper impact, predictive features from the paper’s citation network. In Section \ref{sec:Exp}, we evaluate the proposed approach and compared it with the baseline methods. Section \ref{sec:conclusion} presents the conclusion and future work of the paper.

\section{Related Work}\label{sec:background} 
\par In this section, we present existing work related to the problem of citations count prediction. The current research aimed at various objectives, such as predicting citations for scientific papers \cite{bai2019predicting} 
predicting \textit{h}-index of scientists \cite{mccarty2013predicting}
, and estimating the impact factor of scientific research articles \cite{stallings2013determining}. We can classify the existing research works for the citations count prediction (CCP) based on academic features (Section \ref{sec:Rel_acdFet}), graph of the scientific papers (Section \ref{sec:Rel_graphFet}), post-publication contents of the article (Section \ref{sec:Rel_contFet}), and the hybrid model based CCP (Section \ref{sec:Rel_predMod}).
\subsection{Academic features based CCP} \label{sec:Rel_acdFet}
\par Academic features such as authors' features, link-based features, and posterior features form part of the first category of information used in citations counts. For example, the study \cite{castillo2007estimating} calculates each author's features, including the hub value, the authority's importance, the number of co-authors, the number of articles published and the number of citations. Then they considered the number, average, and maximum value of the features of authors as characteristics of the paper. The authors of this article found that the author's information in an article will help in predicting the published paper's future citations count. The researchers used 20 articles-and journal-based features in another work \cite{lokker2008prediction}, suggesting that citations can also be correctly predicted for two years using data released within three weeks. In \cite{ref13}, researchers used the Latent Dirichlet Allocation (LDA), which reflects the diversity of papers to discover topics and use the entropy of probability distribution over the research topic. Later, in \cite{ref14}, the authors improved earlier work by adding a few new features, such as the novelty of the paper, which is measured as the average of Kullback–Leibler's variance between an article and its references.
\subsection{Graph-based CCP} \label{sec:Rel_graphFet}
\par The graph of research articles contributes to the second type of information used in the citations count prediction. The latest papers in the citation network deal with this problem as a link prediction. In \cite{ref39}, the researchers used the co-authors' empirical network and metric for the author's centrality to predict the highly cited documents. To classify the widely-cited articles, the authors in \cite{ref40} analyzed the document features of the citation network. Moreover, \cite{ref41} built a bipartite network of paper and terms and evaluated the network to categorize publications with the highest impact factor. Later on, Livne et al. \cite{ref15} used a few additional features, namely the \textit{g} index, and six other features specified in \cite{ref18}, to define the citation network's functionality.

\subsection{Content based CCP} \label{sec:Rel_contFet}
\par Post-publishing data for the papers, such as the article's subject area, the references, authors' data, the publishing location (e.g., the conferences or journal), and the article's content were included in the third category of works. Such details are accessible immediately after an article has been published, contributing to understanding its long-term effects. For example, social network information, publication venue, reference list, topics, and the author's information is used in \cite{ref42} to predict if the article is intended to improve \textit{h} indexes for authors within five years of publication. In \cite{sun2009ranking}, the authors examined the impact of the author, publication venue, and content features across various heterogeneous networks. In \cite{ibanez2009predicting}, authors improved citations count prediction performance by incorporating the content of the paper by identifying highly cited keywords in the text.

\subsection{CCP based on hybrid model} \label{sec:Rel_predMod}
Multi-features based citations count predictions have been researched in recent years to improve the performance of the system \cite{robson2016can,sohrabi2017effect}.
These studies use a combination of features such as citation-based features, journal-based features, and authors-based features in the predictive system. The authors used the abstract length, title length, publication year, the prestige of the journal, page number, author's name, and the number of papers in \cite{robson2016can} to predict the article's citations count. Subsequently, the authors proposed works using additional features such as the number of authors, abstract length, title length, and quartile in the SCImago in \cite{sohrabi2017effect}. 
\par In the citations count problem, few articles use more distinctive features and modeling techniques to estimate their citations count. Callaham et al. \cite{callaham2002journal} utilized decision trees to estimate the citations counts of 204 articles of emergency medicine specialty meetings held in 1991. In this paper, authors have considered several nuanced features, including the subjective ``newsworthiness'', the control group, the binary variables reflecting the existence or absence of a specific hypothesis, the number of subjects, and the qualitative scores extracted from the entire study. The authors used several interesting features in \cite{kulkarni2007characteristics}, such as corporate support, group authorship, news media reporting of research, study locations, etc., and applied linear regressions to analyze the citations count of 328 healthcare papers published during 1999 to 2000. The researchers conducted a 4-quartile citations count prediction task in \cite{mcgovern2003exploiting} with a data set of 30199 articles from arXiv.

\par Compared to the existing works, we have considered more important features based on observations of the real world with a citation and co-authorship network. We have also used a Graph Convolutional Network (GCN) based model for estimating the citations count for each published article.

\section{Problem Definition and Dataset}\label{sec:data} 
In this section, first, we present the citations count prediction problem (Section \ref{sec:pr_def}) and then, we describe the dataset used for the analysis (Section \ref{sec:ds_ds}) along with the data preparation in Section \ref{sec:ds_dp}.
\subsection{Problem Definition} \label{sec:pr_def}
Given the literary collection $D$, the citations count ($C_d$) of a research article $d\in D$ can be defined as:
\begin{equation}
    citing(d)=\{d'\in D: d' cites \;d\}
\end{equation}
\begin{equation}
    C_d=|citing(d)|
\end{equation}
Our proposed model takes a number of features, like citation's quality, popularity, author's h-index, and the h-index of the venue 
(detailed discussion is provided in Section \ref{sec:Met_features}) as input and produces the citations count of the research article $d$, which it can possibly receive after the time period $\Delta$.
\begin{figure}[h]
    \centering
    \includegraphics[width=1.0\linewidth,height=30cm,keepaspectratio]{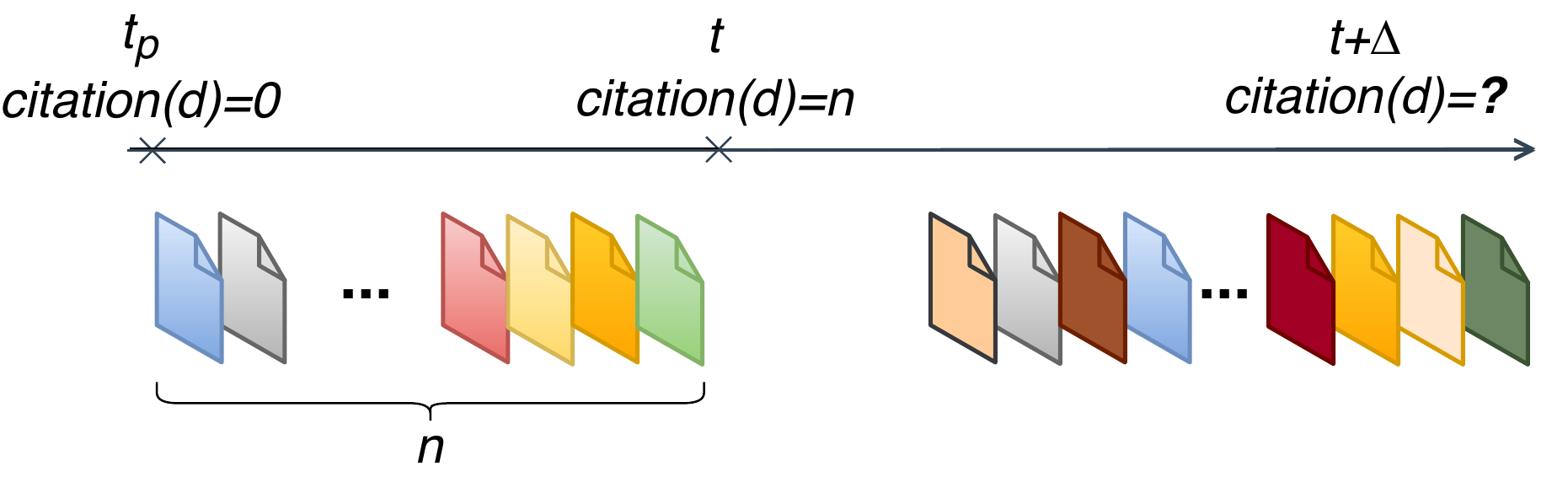}
    \caption{Illustration of citations count prediction.}
    \label{fig:fig2c}
\end{figure}
\par Figure \ref{fig:fig2c} illustrate the citations count prediction problem. Suppose a given publication has received `$n$' number of citations from the date of publications ($t_p$) to the current timestamp $t$, our objective is to develop a model to predict the citations count of a research article $d$ in the future (i.e, after one year, five years, or ten years)

\subsection{Dataset}\label{sec:ds_ds}
\par We conducted our experiments over the real-world academic dataset provided by ArnetMiner (AMiner)\footnote{https://www.aminer.cn/aminernetwork}. It has a large collection of articles published from 1960 to 2011. For our experiment, we have created co-citations network of research articles from the computer science domain, consisting of more than 0.8 million published articles. For each article in this network, we also have information related to publications like the h-index of each author, the number of authors in an article, or the publication venue (see Table \ref{tab:table1}) that can be used as features for the prediction problem of citations counts.
\begin{table}[htbp]
\centering
\caption{Paper information considered in this dataset}
\label{tab:table1}
\resizebox{.45\textwidth}{!}{%
\begin{tabular}{|l|l|}
\hline
\textbf{Terms}      & \textbf{Description}         \\ \hline
Index      & Index id of the paper                  \\ \hline
Title      & Paper title                            \\ \hline
Abstract   & Abstract of the paper                  \\ \hline
Authors    & Authors name listed in the paper       \\ \hline
References & The id of each paper in the references \\ \hline
Venue      & Publication venue                      \\ \hline
Year       & Publication Year                       \\ \hline
\end{tabular}%
}
\end{table}
\subsection{Data Preparation}\label{sec:ds_dp}
\par Upon obtaining the raw data, we carried out the following pre-processing steps. Firstly, we address the missing values of various features like the author's name, affiliations, and publication venues, etc. When the author or publication venue is not present, we delete the row. However, we fill the cell value with a `NaN' if only the affiliation is missing. We also consider articles with no references as some papers might not have attracted citations. Secondly, we remove some anomalous values from the dataset. For example, in a few records, authors have published more than 1000 articles in a single year, which is unrealistic. Thirdly, we normalized the values of each column of features to ensure that the data is consistent. 

\section{Methodology}\label{sec:method}
In this section, we first discuss the factors that can drive a paper's citations count (Section \ref{sec:Met_features}) and then the proposed architecture for estimating the citations count of a research article (Section \ref{sec:Arch}).  
\subsection{Factors driving papers' citations count} \label{sec:Met_features}
We classify various features for predicting the citations counts, into four different categories, namely i) Article's features, ii) Author's features, iii) Venue's features, and iv) Network features. Following, we describe the rationale behind selecting these features.
\begin{enumerate}
\item \par \textbf{Article's features:} The citation quality, popularity and the diversity of papers are considered in the article's features.

\begin{enumerate}
\item \par Citation Quality: Researchers usually select references from articles that are close to their research area. The quality of the reference articles represents the quality of the document. If the article cites the latest hot-spot research topic or articles with more citations, it is more likely that the article will attract more coverage and attract more citations in the coming years \cite{lovaglia1991predicting}. In this article, we use the reference articles' average citation to evaluate the citation quality of the document. Let us assume that, a given article ($d$) has `$r$' number of references, then the citation quality for article $d$ can be computed as:
\begin{equation}
    Citation \; quality(d)= \frac{1}{r} \sum_{i=1}^{r} C_{di}
\end{equation}
where $C_{di}$ is the citations count of article $di \in r$.

\item \par Popularity: We assume that compared to unpopular ones, popular topics receive more attention. Thus, papers dealing with popular topics,  get citations relative easier. 
To capture this effect, we quantify the popularity of each topic $z\in Z$ across the overall corpus $D$ as:
\begin{equation}
    Popularity(z)= \frac{1}{\vert D \vert} \Sigma_{d \in D} p(\frac{z}{\vert d \vert}) \times C_d
\end{equation}
where $p(\frac{z}{\vert d \vert})$ is the probability that paper $d$ distributes on topic $z$, and $c_d$ is the number of citations for article $d$. According to the concept of popularity in \cite{ref31}, we measure an article's average popularity as:
\begin{equation} 
\mathrm{Popularity}(\mathrm{d})=\displaystyle \frac{1}{\vert Z\vert } \Sigma_{z\in Z}popularity(z)\times p(z\vert d) 
\end{equation}
\item \par Diversity: A paper's diversity is described as its topic distribution breadth. The broad range of topics of an article shows a large number of readers and is likely to be highly cited. Using Shannon entropy, we quantify the diversity of the paper. According to the concept of diversity in \cite{ref32}, we measure the average diversity as:
\begin{equation} 
\mathrm{Diversity}(\displaystyle \mathrm{d})= \frac{1}{\vert Z\vert } \Sigma_{z\in Z}-p(z\vert d)\log p(z\vert d) 
\end{equation}
\end{enumerate}

\item \par \textbf{Author's Features:} The features associated with the author are the number of papers published by the first author, highest h-index, total h-indices, average h-index, first author's h-index, average citations count, first author's citations count, highest citations count, the average number of papers published by the authors, the number of papers published by the highest h-index author, and the number of co-author(s).

\begin{enumerate}
\item \par Author's Citation: The number of citations received by the author expresses the scientific capacity of the author. The total citations and average citations reflect each of the author's maximum influence and average influence.

\item \par Author's h-Index: The h-index is a number that reflects the impact and productivity of the authors in the research/academic career \cite{ref33}. Higher is the h-index of an author, the more significant the impact of her/his publications is. The h-index can also relate to the time length spent in the research. It is more favored to the senior researchers who are engaged with many years than early career researchers to measure the research field's impact.

\item \par Author's Ability: The author's abilities involve the number of co-authors and productivity, which can be defined as follows:
\begin{enumerate}
\item The number of co-authors represents the author's professional competence with peers. People tend to cite the papers with relatively higher number of co-authors \cite{ref34}.
\item An author's productivity depicts the results of its research. The more papers an individual publishes, intuitively, the higher the average citation S/he will receive in the future.
\end{enumerate}
\end{enumerate}

\item \par \textbf{Venue Features:} The impact factor of the journal in which the article has been published is found to be a relevant attribute \cite{callaham2002journal}. Thus,  the journal wherein the article is published can also help in predicting the citations count. Factors that come under the venue features are databases indexing, publishing venue (Journal or conferences), venue rank, h-index, and the average citation of the publication venue. 

\item \par \textbf{Network Features:} We created a directed co-citation network of the published articles, where the vertices represents documents and edges form the cited connection between them. The out-degrees calculate how many times a document is cited and in-degrees refers to the number of references appear in the document.
\end{enumerate}

\subsection{Proposed Architecture}\label{sec:Arch}
We have proposed a Graph Convolutional Neural Network (GCN) \cite{kipf2016graph} based framework to address the citation count prediction problem. GCN has recently gained a lot of attention since the various graph-structured data \cite{ref26} have been successfully implemented. The co-citation graph can be described as $G = (V, E)$ in which $V$ denotes the set of articles, and $e_{ij} \in E$ represents an edge between two articles $v_i$ and $v_j$ ($v_i$,$v_j$ $\in$ $V$) if an article $v_i$ cites $v_j$. In addition, the graph is illustrated using the following:
\begin{itemize}
    \item In this graph, each node $v \in V$ has a feature vector $F_v$ of size $m$. So we have $X: n \times m$ feature matrix for $n$ nodes.
    \item The adjacency matrix $M$, representing the graph structure, is also a key element.
    \item The propagation rule will generate a node-level output $O: n \times F$, where $F$ represents feature vector of each output node.
    \item Each neural network layer is defined as:
   \begin{equation}
        F^{k+1}= f(F^k, M)
    \end{equation}
\end{itemize}
where $F^{k+1}$ represents the hidden layer node matrix at $(k+1)^{th}$ level and it is equivalent to the function of previous hidden layer node matrix $F^k$ at $k^{th}$ level and the adjacency matrix $M$. $F$ can be taken as the feature matrix $X$ at initial level, i.e $F^{0} = X$ and $O$ at final level. $O$ represents the graph level output.
\par
\begin{figure*}[htbp]
    \centering
    \includegraphics[width=0.9\linewidth,height=25cm,keepaspectratio]{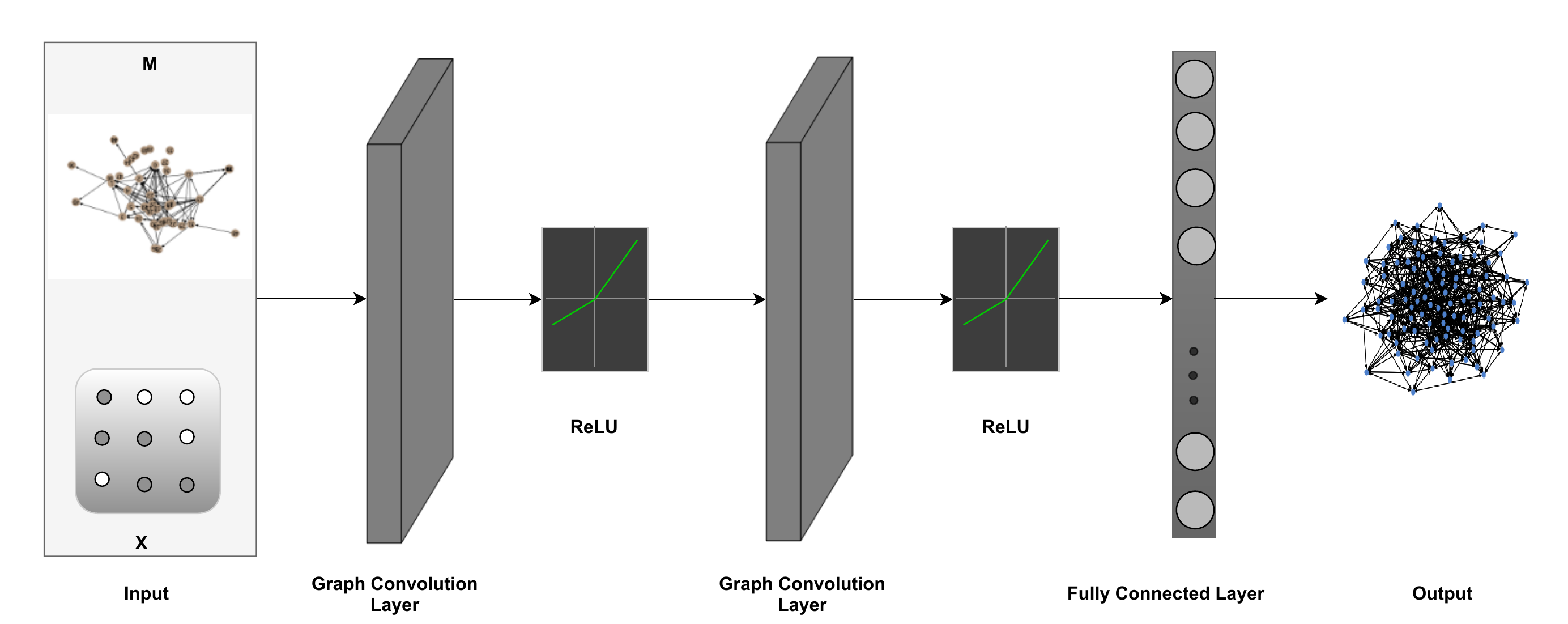}
    \caption{A graph convolutional network architecture for predicting the citations count of the article.
    }
    \label{fig:fig3}
\end{figure*}

 In particular, we design a GCN-based predictive model with two layers of graph convolution, an input layer and an output layer. The architecture of the proposed predictive model is shown in Figure \ref{fig:fig3}. The proposed model uses a node link adjacency matrix ($M$) and a node feature matrix ($X$) as input, where the node-level features are author features, content features, network features, etc. (discussed in Section \ref{sec:Met_features}). It takes the topological data from the graphs and generates feature vectors through a GCN that contains all the features of the topological node. During training, the model is updated at each stage, and an adaptive learning rate optimizer algorithm called ``Adam'' \cite{kingma2014adam} optimizes the model.
\par The GCN uses the data flow between edges in the graph to create a graph embedding. Once we create an embedded graph influenced by all neighbors, we can consider the entire graph as one graph. Here we assume that the number of citations received in a particular paper is related to the number of citations received by its neighbor. The relationship between documents is displayed with an adjacency matrix $M$. With our notation, we can define a GCN layer as:
\begin{equation}
    f(F^k, M)= \sigma (MF^kW^k)
\end{equation}
where $F^k$ is $k^{th}$ layer node-level features; $W^k$ represents layer-specific trainable weight matrix; $M$ shows the network adjacency matrix; $\sigma$ is a non-linear activation function like $ReLU(\cdot)$; and $F^{k+1}$ is the output of the GCN layer. The issue with this model is that it does not normalize the adjacency matrix. In this case, the adjacency matrix and feature matrix multiplication will create an entirely different feature space. This can be solved by performing normalization of the matrix, which can be accomplished by introduction of an additional inverse diagonal node degree matrix $m$, such that the rows of $m^{-1}M$ sums to 1. With this, the multiplication becomes more similar to taking the average of adjacent nodes. This lead to symmetric normalization i.e, $m^{-1/2}Mm^{-1/2}$,  and it more than just a mere averaging of neighboring features. Another problem is that the computation does not consider the self-features of node itself. This problem can be solved by taking the support of the identity matrix $I$.
In this work, the above methods are combined to be used in a propagation rule, and the final layer-wise  propagation rule is defined as:
\begin{equation}
    f(F^k, M)= \sigma (\hat{m}^{-1/2}\hat{M}\hat{m}^{-1/2}F^kW^k)
\end{equation}
where, $\hat{M} = M + I$; with $I$ defined as identity matrix and $\hat{m}$ is the diagonal degree node matrix of $\hat{M}$.

\section{Experimental Results and Analysis}\label{sec:Exp}
This section discusses the experimental setup, evaluation metrics, 
and a comparative analysis with the baseline predictive models.
\subsection{Experimental Setup}\label{sec:ExpSetUp}
\par We compared our experimental results with four baseline algorithms: Linear Regression (LR) \cite{lovaglia1991predicting}, Random Forest (RF) \cite{ref37}, XGBoost \cite{ref36}, and Deep Neural Networks (DNN) \cite{ref38}. The parameter values chosen in our experiment for the baseline algorithms are shown in Table \ref{tab:table3_1}.
In this paper, we cover three different cases for predicting the citations count of the research articles. First, with the dataset of research articles published from January 2010 to December 2010 for 1-year citations count prediction, second with the research articles published from January 2006 to December 2006 for 5-years citations count prediction, and third with the research articles published from January 2010 to December 2010 for 10-years citations count prediction. For all citations count prediction the details of the dataset chosen for training and testing purposes and the temporal duration in each case are given in Table \ref{tab:table1_11}.  To demonstrate the prediction model's performance and to achieve a more stable and reliable model, we use a 10-fold cross-validation method to carry out experiments.
\begin{table}[htbp]
\centering
\caption{Parameters settings in the baseline models}
\label{tab:table3_1}
\resizebox{.48\textwidth}{!}{%
\begin{tabular}{|l|l|l|}
\hline
\textbf{Models} & \textbf{Parameters}& \textbf{Values}     \\ \hline
Linear Regression       & - & -     \\ \hline
Random Forest           & \begin{tabular}[c]{@{}l@{}}n estimators\\ max\_depth\end{tabular} & \begin{tabular}[c]{@{}l@{}}500\\ 2\end{tabular}                             \\ \hline
XGBoost                 & \begin{tabular}[c]{@{}l@{}}learning\_rate\\min\_samples\_split\\max\_depth\\n\_estimators \end{tabular} & \begin{tabular}[c]{@{}l@{}}0.001\\2\\4\\500 \end{tabular} \\ \hline
Deep Neural Networks & \begin{tabular}[c]{@{}l@{}}Batch size\\ Learning rate\\ Number of Hidden units\\ Dropout rate\\ Epochs\\ Optimizer\\ Activation function\end{tabular} & \begin{tabular}[c]{@{}l@{}}256\\ 0.001\\ 512\\ 0.2\\ 200\\Adam \\ ReLu\end{tabular} \\ \hline
\end{tabular}%
}
\end{table}
\begin{table*}[htbp]
\centering
\large
\caption{Temporal Split for 3 cases of the dataset} 
\label{tab:table1_11}
\resizebox{.80\textwidth}{!}{%
\begin{tabular}{|l|l|l|l|l|}
\hline
\textbf{Cases} & \textbf{Training samples} & \textbf{Testing samples} & \textbf{Training period time}& \textbf{Temporal duration $(\Delta)$}     \\ \hline
Case 1 & 3,55,500  & 39,500 & Jan 2010 - Dec 2010 & 1-Year citations count prediction \\ \hline
Case 2 & 2,88,000 & 32,000 & Jan 2006 - Dec 2006 & 5-Years citations count prediction \\ \hline
Case 3 & 1,26,000 &14,000 & Jan 2001 - Dec 2001 & 10-Years citations count prediction \\ \hline
\end{tabular}%
}
\end{table*}
%\vspace{-0.3cm}
\subsection{Evaluation Metrics}
This section presents evaluation metrics such as MAE, RMSE, and MAPE to measure the model's prediction accuracy. These error metrics represent the difference between the predicted values and the real values, and the smaller the difference, the better the model's performance. These error metrics are formulated as:
\begin{equation}
    MAE= \frac{1}{n} \sum_{i=1}^{n}\mid y_i-{y'_i}\mid
\end{equation}
\begin{equation}
    RMSE= \sqrt{\frac{1}{n} \sum_{i=1}^{n}(y_i-{y'_i})^2}
\end{equation}
\begin{equation}
    MAPE= \frac{1}{n} \sum_{i=1}^{n}\mid \frac{y_i-{y'_i}}{y_i} \mid
\end{equation}
Further, we considered two widely used metrics in regression tasks: coefficient of determination ($R^2$) and adjusted $R^2$ to evaluate the model's prediction performance quantitatively. $R^2$ is used in the context of statistical models whose aim is to predict future outcomes based on related features. The value of $R^2$ lies in the range of 0 to 1. A larger $R^2$ shows better performance of the model. It can be computed as:
\begin{equation} 
R^{2}=1-\displaystyle \frac{\Sigma(y_{i}-{y'}_{i})^{2}}{\Sigma(y_{i}-\bar{Y})^{2}}
\end{equation}
where `$y_i$', `$y'_i$' represents the actual and predicted values, and `$\bar{Y}$' is the average value of the samples.
\par A baseline-equivalent model, will provide $R^2$ with 0. Higher the $R^2$ value, better the performance of the model. The optimal model will give $R^2$ equals to 1 with all the correct predictions. However, the $R^2$ value either increases or remains the same when introducing new features to a model. For adding new features that do not add any value to the model, $R^2$ is not penalized. Therefore, the adjusted $R^2$ is an enhanced version of the $R^2$ metric, computed as:
\begin{equation} 
Adjusted \;R^{2}=1-\displaystyle \frac{(1-R^2)(n-1)}{n-(P+1)}
\end{equation}
where `$n$' represents the number of samples, `$R^2$' is the coefficient of determination, and `$P$' is the number of predictors.

\subsection{Comparative Analysis}
To test the effectiveness of the GCN-based citations count prediction model, the prediction performance of GCN is compared against four baseline models: Linear Regression (LR), Random Forest (RF), XGBoost, and the Deep Neural Networks. The comparison is made in terms of MAE, RMSE, MAPE, $R^2$, and the Adjusted $R^2$. The citations count prediction performances of the baseline algorithms and the GCN-based proposed model are shown in Table \ref{tab:table2_1}, \ref{tab:table2_5}, and \ref{tab:table2_10} for 1-year, 5-years and 10-years time frame respectively. In general GCN-based model has shown better performance compared to the baseline models, especially for 5-years and 10-years cases. For the 1-year case, the GCN-based model has shown less error for MAE metric however, the values are slightly higher (relative to Deep Neural Network model) for RMSE and MAPE. 
\par From Tables \ref{tab:table2_1}, \ref{tab:table2_5}, and \ref{tab:table2_10}, we observed the following points.
\begin{itemize}
    \item The GCN-based proposed model outperforms all baseline models, across all the years in terms of MAE. This is especially the case for long-term cases (that is 5-years and 10-years cases).
    We also observe that the MAE values of all predictive models decreases as the predictive period ($ \Delta $) increases. 
    \item Unlike MAE, the RMSE value of Deep Neural Network model is less than other predictive models for short-term citations count (1-year). For long-term citation, the GCN-based model yields lower RMSE value than other baseline models. Similar to MAE, RMSE value also decreases as predictive duration increases for all the predictive models.
    \item We observed that the MAPE values for all the predictive models are mixed. The GCN-based model yields lower MAPE value than other models for a 5-years and 10-years citations count, which shows that the GCN-based model performs well for the long-term prediction, but it performs slightly worse for the short-term than XGBoost and Deep Neural Network-based models for RMSE and MAPE metrics.
    \item The proposed model has better prediction performance in terms of $R^2$ and adjusted $R^2$ than the other four baseline models. For 1-year citations count prediction, $R^2$ value of the GCN-based model is increased by 15.78\% for the LR, 7.79\% for the RF, 2.94\% for the XGBoost, and 0.66\% for the Deep Neural Networks. $R^2$ value of the GCN-based model for 5-years citations count prediction is increased by 22.58\%, 16.80\%, 9.50\%, and 5.23\% for LR, RF, XGBoost, and Deep Neural Networks respectively. These results indicate that the proposed GCN-based predictive model is more robust in terms of performance compared to other baseline models, specifically for long-term prediction compared to short-term prediction.
\end{itemize}
\par From these results, we notice that non-linear models perform well compared to the linear model. It can be noted that the prediction performance of all models is better for 10-years citations count prediction compared to 5-years, which is further better than for the 1-year case, in terms of both the error rate and $R^2$ values. This is possibly due to the fact that, in a 10-years case, relatively more reliable information is available in the year 2010 for articles (and their corresponding authors) which got published in 2001 compared to the case where the articles got published in 2010 (1-year case) for predicting the future citations count of the article. 
\par The conventional neural networks like DNN and CNN cannot properly handle the graph input data as they end up stacking the node's features in a non-useful (for example, random) fashion. However, GCN propagates each node irrespective of the order of nodes. In graph analysis, the dependency information between the nodes are represented by edges, and the performance depends on the edges' information. But, in the conventional neural networks, that information is considered as the node features.
\vspace{-0.3cm}

\begin{table}[htbp]
\centering
\caption{The performance of various machine learning techniques for 1-Year citations count prediction}
\label{tab:table2_1}
\resizebox{.45\textwidth}{!}{
\begin{tabular}{|c|c|c|c|c|c|}
\hline
Algorithm         & MAE               & RMSE             & MAPE             & R$^2$       & Adj R$^2$  \\ \hline
Linear Regression & 11.8531 &41.2517  &0.4814  &0.6189  &0.5617 \\ \hline
Random Forest     & 9.6428  &38.8126  &0.3575  &0.6648  &0.6145 \\ \hline
XGBoost           &  9.1179 &31.6872  &\textbf{0.2761}  &0.6961  &0.6505 \\ \hline
Deep Neural Network    &  8.5219 &\textbf{24.4264}  &0.2857  &0.7119  &0.6686 \\ \hline
GCN  &  \textbf{8.4335} &28.6757  &0.2974  &\textbf{0.7166}  &\textbf{0.6740} \\ \hline
\end{tabular}
}
\end{table}
\vspace{-0.7cm}
\begin{table}[htbp]
\centering
\caption{The performance of various machine learning techniques for 5-Years citations count prediction}
\label{tab:table2_5}
\resizebox{.45\textwidth}{!}{
\begin{tabular}{|c|c|c|c|c|c|}
\hline
Algorithm         & MAE               & RMSE             & MAPE             & R$^2$       & Adj R$^2$  \\ \hline
Linear Regression & 5.8817 &11.8526  &0.3604  &0.6984  &0.6531   \\ \hline
Random Forest     & 5.4170 &11.1250  &0.3125  &0.7329  &0.6928  \\ \hline
XGBoost           & 4.7816 &10.6129  &0.2375  &0.7818  &0.7490  \\ \hline
Deep Neural Network    & 4.6638 &9.8976   &0.2219  &0.8135  &0.7855 \\ \hline
GCN  & \textbf{4.1214} &\textbf{9.1248}   &\textbf{0.1883}  &\textbf{0.8561}  &\textbf{0.8345}  \\ \hline 
\end{tabular}
}
\end{table}
\vspace{-0.7cm}
\begin{table}[htbp]
\centering
\caption{The performance of various machine learning techniques for 10-Years citations count prediction.}
\label{tab:table2_10}
\resizebox{.45\textwidth}{!}{
\begin{tabular}{|c|c|c|c|c|c|}
\hline
Algorithm         & MAE               & RMSE             & MAPE             & R$^2$       & Adj R$^2$  \\ \hline
Linear Regression & 3.5719 &4.6252  &0.2111  &0.7252  &0.6839   \\ \hline
Random Forest     & 2.9986 &3.1779  &0.1964  &0.7638  &0.7283  \\ \hline
XGBoost           & 2.6310 &2.8353  &0.1615  &0.8516  &0.8293  \\ \hline
Deep Neural Network    & 2.4819 &2.4510  &0.134   &0.8821  &0.8644 \\ \hline
GCN  & \textbf{1.6518} &\textbf{2.1654}  &\textbf{0.1072}  &\textbf{0.9157}  &\textbf{0.9030}  \\ \hline
\end{tabular}
}
\end{table}
\section{Conclusion and Future Scope}\label{sec:conclusion} 
\par In this work, we studied the citations count prediction as a regression problem, for predicting the citations count of research articles for a given period. In particular, we explored the graph convolutional network-based model, which takes two matrices. The first is the adjacency matrix of the  co-cited articles graph. The second one is the feature matrix consisting of various features for the articles. This way GCN can capture the relationships among the co-cited articles efficiently. We evaluated GCN based method (along with baseline methods) on a co-citation network of more than 0.8 million published articles from the Computer Science domain. Our experimental results indicated that the GCN-based predictive model performed better compared to baseline methods. 
In particular, for 10-years and 5-years citations count prediction cases, the model achieved the best score in terms of error rates. In addition, the $R^2$ value achieved by the GCN-based model has the best value indicating that the GCN model is more robust compared to baseline models. We plan to extend this work in many different ways. In this work, we selected articles only from the computer science domain. However, for future work, we would like to increase the dataset by selecting articles from different domains. In addition, we would also like to explore other sources of datasets, such as DBLP to increase the volume of articles. 
\vspace{-0.3cm}
\section*{Acknowledgment}
This research is financially supported by DORA Plus grant and H2020 SoBigData++ project, and CHIST-ERA project SAI.
\bibliography{ref}

\begin{thebibliography}{10}

\bibitem{ref20}
Barbara~J Robson and Aur{\'e}lie Mousqu{\`e}s.
\newblock Can we predict citation counts of environmental modelling papers?
  fourteen bibliographic and categorical variables predict less than 30\% of
  the variability in citation counts.
\newblock {\em Environmental Modelling \& Software}, 75:94--104, 2016.

\bibitem{thelwall2018could}
Mike Thelwall and Tamara Nevill.
\newblock Could scientists use altmetric. com scores to predict longer term
  citation counts?
\newblock {\em Journal of informetrics}, 12(1):237--248, 2018.

\bibitem{ref21}
Babak Sohrabi and Hamideh Iraj.
\newblock The effect of keyword repetition in abstract and keyword frequency
  per journal in predicting citation counts.
\newblock {\em Scientometrics}, 110(1):243--251, 2017.

\bibitem{ref8}
Emre Sarig{\"o}l, Ren{\'e} Pfitzner, Ingo Scholtes, Antonios Garas, and Frank
  Schweitzer.
\newblock Predicting scientific success based on coauthorship networks.
\newblock {\em EPJ Data Science}, 3(1):9, 2014.

\bibitem{ref4}
Xuanyu Cao, Yan Chen, and KJ~Ray Liu.
\newblock A data analytic approach to quantifying scientific impact.
\newblock {\em Journal of Informetrics}, 10(2):471--484, 2016.

\bibitem{singh2017understanding}
Mayank Singh, Ajay Jaiswal, Priya Shree, Arindam Pal, Animesh Mukherjee, and
  Pawan Goyal.
\newblock Understanding the impact of early citers on long-term scientific
  impact.
\newblock In {\em 2017 ACM/IEEE Joint Conference on Digital Libraries (JCDL)},
  pages 1--10. IEEE, 2017.

\bibitem{newman2009first}
Mark~EJ Newman.
\newblock The first-mover advantage in scientific publication.
\newblock {\em EPL (Europhysics Letters)}, 86(6):68001, 2009.

\bibitem{xiao2016modeling}
Shuai Xiao, Junchi Yan, Changsheng Li, Bo~Jin, Xiangfeng Wang, Xiaokang Yang,
  Stephen~M Chu, and Hongyuan Zha.
\newblock On modeling and predicting individual paper citation count over time.
\newblock In {\em IJCAI}, pages 2676--2682, 2016.

\bibitem{kipf2016graph}
Thomas Kipf.
\newblock Graph convolutional networks, 2016.

\bibitem{gilmer2017neural}
Justin Gilmer, Samuel~S Schoenholz, Patrick~F Riley, Oriol Vinyals, and
  George~E Dahl.
\newblock Neural message passing for quantum chemistry.
\newblock {\em arXiv preprint arXiv:1704.01212}, 2017.

\bibitem{bruna2017community}
Joan Bruna and X~Li.
\newblock Community detection with graph neural networks.
\newblock {\em Stat}, 1050:27, 2017.

\bibitem{chen2018gc}
Jinyin Chen, Xuanheng Xu, Yangyang Wu, and Haibin Zheng.
\newblock Gc-lstm: Graph convolution embedded lstm for dynamic link prediction.
\newblock {\em arXiv preprint arXiv:1812.04206}, 2018.

\bibitem{vijayan2018forecasting}
Raghavendran Vijayan and George Mohler.
\newblock Forecasting retweet count during elections using graph convolution
  neural networks.
\newblock In {\em 2018 IEEE 5th International Conference on Data Science and
  Advanced Analytics (DSAA)}, pages 256--262. IEEE, 2018.

\bibitem{lovaglia1991predicting}
Michael~J Lovaglia.
\newblock Predicting citations to journal articles: The ideal number of
  references.
\newblock {\em The American Sociologist}, 22(1):49--64, 1991.

\bibitem{ref37}
G~James et~al.
\newblock An introduction to statistical learning springer new york.
\newblock {\em New York, NY}, 2013.

\bibitem{ref36}
Tianqi Chen and Carlos Guestrin.
\newblock Xgboost: A scalable tree boosting system.
\newblock In {\em Proceedings of the 22nd acm sigkdd international conference
  on knowledge discovery and data mining}, pages 785--794, 2016.

\bibitem{ref38}
Ali Abrishami and Sadegh Aliakbary.
\newblock Predicting citation counts based on deep neural network learning
  techniques.
\newblock {\em Journal of Informetrics}, 13(2):485--499, 2019.

\bibitem{bai2019predicting}
Xiaomei Bai, Fuli Zhang, and Ivan Lee.
\newblock Predicting the citations of scholarly paper.
\newblock {\em Journal of Informetrics}, 13(1):407--418, 2019.

\bibitem{mccarty2013predicting}
Christopher McCarty, James~W Jawitz, Allison Hopkins, and Alex Goldman.
\newblock Predicting author h-index using characteristics of the co-author
  network.
\newblock {\em Scientometrics}, 96(2):467--483, 2013.

\bibitem{stallings2013determining}
Jonathan Stallings, Eric Vance, Jiansheng Yang, Michael~W Vannier, Jimin Liang,
  Liaojun Pang, Liang Dai, Ivan Ye, and Ge~Wang.
\newblock Determining scientific impact using a collaboration index.
\newblock {\em Proceedings of the National Academy of Sciences},
  110(24):9680--9685, 2013.

\bibitem{castillo2007estimating}
Carlos Castillo, Debora Donato, and Aristides Gionis.
\newblock Estimating number of citations using author reputation.
\newblock In {\em International Symposium on String Processing and Information
  Retrieval}, pages 107--117. Springer, 2007.

\bibitem{lokker2008prediction}
Cynthia Lokker, K~Ann McKibbon, R~James McKinlay, Nancy~L Wilczynski, and
  R~Brian Haynes.
\newblock Prediction of citation counts for clinical articles at two years
  using data available within three weeks of publication: retrospective cohort
  study.
\newblock {\em Bmj}, 336(7645):655--657, 2008.

\bibitem{ref13}
Rui Yan, Jie Tang, Xiaobing Liu, Dongdong Shan, and Xiaoming Li.
\newblock Citation count prediction: learning to estimate future citations for
  literature.
\newblock In {\em Proceedings of the 20th ACM international conference on
  Information and knowledge management}, pages 1247--1252, 2011.

\bibitem{ref14}
Rui Yan, Congrui Huang, Jie Tang, Yan Zhang, and Xiaoming Li.
\newblock To better stand on the shoulder of giants.
\newblock In {\em Proceedings of the 12th ACM/IEEE-CS joint conference on
  Digital Libraries}, pages 51--60, 2012.

\bibitem{ref39}
Emre Sarig{\"o}l, Ren{\'e} Pfitzner, Ingo Scholtes, Antonios Garas, and Frank
  Schweitzer.
\newblock Predicting scientific success based on coauthorship networks.
\newblock {\em EPJ Data Science}, 3(1):9, 2014.

\bibitem{ref40}
Daniel Mcnamara, Paul Wong, Peter Christen, and Kee~Siong Ng.
\newblock Predicting high impact academic papers using citation network
  features.
\newblock In {\em Pacific-Asia Conference on Knowledge Discovery and Data
  Mining}, pages 14--25. Springer, 2013.

\bibitem{ref41}
Peter Klimek, Aleksandar~S Jovanovic, Rainer Egloff, and Reto Schneider.
\newblock Successful fish go with the flow: citation impact prediction based on
  centrality measures for term--document networks.
\newblock {\em Scientometrics}, 107(3):1265--1282, 2016.

\bibitem{ref15}
Avishay Livne, Eytan Adar, Jaime Teevan, and Susan Dumais.
\newblock Predicting citation counts using text and graph mining.
\newblock In {\em Proc. the iConference 2013 Workshop on Computational
  Scientometrics: Theory and Applications}, 2013.

\bibitem{ref18}
Xiaolin Shi, Jure Leskovec, and Daniel~A McFarland.
\newblock Citing for high impact.
\newblock In {\em Proceedings of the 10th annual joint conference on Digital
  libraries}, pages 49--58, 2010.

\bibitem{ref42}
Yuxiao Dong, Reid~A Johnson, and Nitesh~V Chawla.
\newblock Can scientific impact be predicted?
\newblock {\em IEEE Transactions on Big Data}, 2(1):18--30, 2016.

\bibitem{sun2009ranking}
Yizhou Sun, Yintao Yu, and Jiawei Han.
\newblock Ranking-based clustering of heterogeneous information networks with
  star network schema.
\newblock In {\em Proceedings of the 15th ACM SIGKDD international conference
  on Knowledge discovery and data mining}, pages 797--806, 2009.

\bibitem{ibanez2009predicting}
Alfonso Ib{\'a}{\~n}ez, Pedro Larra{\~n}aga, and Concha Bielza.
\newblock Predicting citation count of bioinformatics papers within four years
  of publication.
\newblock {\em Bioinformatics}, 25(24):3303--3309, 2009.

\bibitem{robson2016can}
Barbara~J Robson and Aur{\'e}lie Mousqu{\`e}s.
\newblock Can we predict citation counts of environmental modelling papers?
  fourteen bibliographic and categorical variables predict less than 30\% of
  the variability in citation counts.
\newblock {\em Environmental Modelling \& Software}, 75:94--104, 2016.

\bibitem{sohrabi2017effect}
Babak Sohrabi and Hamideh Iraj.
\newblock The effect of keyword repetition in abstract and keyword frequency
  per journal in predicting citation counts.
\newblock {\em Scientometrics}, 110(1):243--251, 2017.

\bibitem{callaham2002journal}
Michael Callaham, Robert~L Wears, and Ellen Weber.
\newblock Journal prestige, publication bias, and other characteristics
  associated with citation of published studies in peer-reviewed journals.
\newblock {\em Jama}, 287(21):2847--2850, 2002.

\bibitem{kulkarni2007characteristics}
Abhaya~V Kulkarni, Jason~W Busse, and Iffat Shams.
\newblock Characteristics associated with citation rate of the medical
  literature.
\newblock {\em PloS one}, 2(5):e403, 2007.

\bibitem{mcgovern2003exploiting}
Amy McGovern, Lisa Friedland, Michael Hay, Brian Gallagher, Andrew Fast,
  Jennifer Neville, and David Jensen.
\newblock Exploiting relational structure to understand publication patterns in
  high-energy physics.
\newblock {\em Acm Sigkdd Explorations Newsletter}, 5(2):165--172, 2003.

\bibitem{ref31}
Yuxiao Dong, Reid~A Johnson, and Nitesh~V Chawla.
\newblock Can scientific impact be predicted?
\newblock {\em IEEE Transactions on Big Data}, 2(1):18--30, 2016.

\bibitem{ref32}
Rui Yan, Congrui Huang, Jie Tang, Yan Zhang, and Xiaoming Li.
\newblock To better stand on the shoulder of giants.
\newblock In {\em Proceedings of the 12th ACM/IEEE-CS joint conference on
  Digital Libraries}, pages 51--60, 2012.

\bibitem{ref33}
Jorge~E Hirsch.
\newblock An index to quantify an individual's scientific research output.
\newblock {\em Proceedings of the National academy of Sciences},
  102(46):16569--16572, 2005.

\bibitem{ref34}
Steven Bethard and Dan Jurafsky.
\newblock Who should i cite: learning literature search models from citation
  behavior.
\newblock In {\em Proceedings of the 19th ACM international conference on
  Information and knowledge management}, pages 609--618, 2010.

\bibitem{ref26}
Thomas~N Kipf and Max Welling.
\newblock Semi-supervised classification with graph convolutional networks.
\newblock {\em arXiv preprint arXiv:1609.02907}, 2016.

\bibitem{kingma2014adam}
Diederik~P Kingma and Jimmy Ba.
\newblock Adam: A method for stochastic optimization.
\newblock {\em arXiv preprint arXiv:1412.6980}, 2014.

\end{thebibliography}
\bibliographystyle{unsrt}
\end{document}